\documentclass[prd,twocolumn,nofootinbib]{revtex4}
\usepackage{graphicx}
\usepackage{latexsym}
\vspace{2cm}
\def\beq{\begin{equation}}
\def\eeq{\end{equation}}
\def\beqn{\begin{eqnarray}}
\def\eeqn{\end{eqnarray}}

\def\no{\nonumber}
\begin{document}
\title{WMAP constraint on the P-term inflationary model}
\author{Bin Wang}
\affiliation{
Department of Physics, Fudan University, 200433 Shanghai}
\author{Chi-Yong Lin}
\affiliation{
Department of Physics, National Dong Hwa University, Shoufeng, 974 Hualien} 
\author{Elcio Abdalla}
\affiliation{
Instituto de Fisica, Universidade de Sao Paulo,
  C.P.66.318, CEP 05315-970, Sao Paulo} 

\begin{abstract}
In light of WMAP results, we examine the observational constraint on the
P-term inflation. With the tunable parameter $f$, P-term inflation
contains richer physics than D-term and F-term inflationary models. We
find the logarithmic derivative spectral index with $n>1$ on large scales
and $n<1$ on small scales in agreement to observation. We obtained a 
reasonable range for the choice of the gauge coupling constant $g$ in
order to meet the requirements of WMAP observation and the expected number
of the e-foldings. Although tuning $f$ and $g$ we can have larger values 
for the logarithmic derivative of the spectral index, it is not possible 
to satisfy all observational requirements for both, the spectral index and
its logarithmic derivative at the same time.\\ 
PACS number(s): 98.80.Cq
\end{abstract}
\maketitle

\section{Introduction
\label{sec:1}}
One of the fundamental ideas of modern cosmology is that in its very early
epoch of history our universe was dominated by the potential energy of a
slowly rolling scalar field \cite{1}\cite{2}. During such an era the scale
factor grew exponentially in a very short period of time, leading to the
homogeneity and isotropy of the observed universe to a high accuracy. The
inflationary scenario has been supported by COBE \cite{3} and other large
scale galaxy surveys \cite{4}. The recently released first year high
precision data of the WMAP further confirmed the inflationary mechanism
\cite{5}. 

There are various single-field inflationary models. Among them,
non-symmetric grand unified theories which give rise to the inflationary
scenario were constructed more than a decade ago. Besides the standard
model, supersymmetry has been considered both as a blessing and as a curse
for inflationary model building. It is a blessing, primarily because it
allows one to have very flat potential, as well as to fine-tune any
parameters at the tree level. Moreover it seems more natural than the
non-symmetric theories. It is a curse, because during inflation one needs
to consider supergravity, where usually all scalar fields have too big
masses to support inflation. However it was shown that in the $N=1$
generic D-term inflation the inflaton field has superpotentials which
vanish together with their first derivatives and does not acquire a
Hubble scale mass term from the supergravity corrections \cite{6}. It
avoids the general problem of inflation in supergravity. In addition it
was found that by properly choosing the minimal K\"ahler potential, the
$N=1$ supersymmetric F-term inflation does not acquire the large mass term
usually needed in supergravity either \cite{7}.

Recently a new version of hybrid inflation, the ``P-term inflation'' has
been introduced in the context of $N=2$ supersymmetry \cite{8}. It is
intriguing that once one breaks $N=2$ supersymmetry and implements the
P-term inflation in $N=1$ supergravity, this scenario simultaneously leads
to a new class of inflationary models, which interpolates between D-term
and F-term models. 

It is of interest to use the WMAP data to discriminate among the various
single-field inflationary models \cite{kinney}. For single-field inflation
models, the 
relevant parameter space for distinguishing among models is defined by the
scalar index $n$ and the logarithmic derivative of the scalar spectral
index $dn/d(\ln k)$. In D-term inflation it was found that the spectrum is
exactly flat with $n=1$ \cite{8}. In the F-term inflation a small
dependence of the spectral index on momenta was found, which is
qualitatively consistent with the WMAP observation \cite{5}. However, the
variation of the spectral index in F-term inflation was claimed to be too
mild compared to the observational data \cite{9}\cite{10}. It was argued
that this is due to the small Yukawa coupling constant required for
sufficient inflation \cite{10}. The motivation of the present paper is to
use the WMAP data to restrain the P-term inflation parameter. In P-term
inflation the contribution from supergravity has an additional parameter
$f$ in the range $0<f<1$. We would like to explore how the inflationary
regime and its properties depend on the value of the parameter $f$ to meet
observation. 

\section{P-term inflation model
\label{sec:2}}
We first review the P-term inflationary model of Kallosh and Linde
\cite{8}. The action of the $N=2$ supersymmetry is 
\beq \label{eq-1}             
L_{N=2}=L_{s,c}-g\vec{P}\vec{\zeta}
\eeq
where $L_{s,c}$ is the bosonic part of the superconformal action
consisting of the vector multiplet and a charged hypermultiplet. The
second term in (\ref{eq-1}) is the Fayet-Iliopulos term. The potential of
the P-term model can be given by using the symmetries of the theory in a
form suitable for the $N=1$ notation,
\beqn     
V_{N=2} & = & 2g^2(\vert S\Phi_+\vert ^2+ \vert S\Phi_-\vert ^2+\vert
\Phi_+\Phi_- -\xi_+/2\vert ^2) \no \\ 
& + & g^2/2 (\vert \Phi_+\vert ^2 -\vert \Phi_-\vert ^2-\xi_3)^2 \label{eq-2}
\eeqn
where $S$ is for the neutral scalar, $\Phi_+ (\Phi_-)$ for positively
(negatively) charged scalar. $\xi$ is the triplet with $\xi=\sqrt{\vert
  \vec{\xi}\vert^2}=\sqrt{\xi_+\xi_- +(\xi_3)^2}, \xi_{\pm}=\xi_1
+i\xi_2$. Here $g$ is the gauge coupling constant. 

The P-term potential corresponds to that of an $N=1$ model
\beq \label{eq-3} 
V=\vert \partial W\vert^2+g^2 D^2/2
\eeq
with a superpotential and a D-term given by
\beq \label{eq-4} 
W=\sqrt{2}gS(\Phi_+\Phi_- -\xi_+/2), \quad
D=\vert\Phi_+\vert^2-\vert\Phi_-\vert^2-\xi_3. 
\eeq
It coincides with the D-term inflation when $\xi_+=\xi_-=0$,
$\xi_3=\vert\vec{\xi}\vert$ and $\lambda=\sqrt{2}g$. On the other hand
choosing $\xi_+=\xi_-=\xi$ and $\lambda=\sqrt{2}g$, we can recover the
F-term inflation model. 

$N=2$ supersymmetry can be coupled to $N=1$ supergravity by choosing the
minimal K$\ddot{a}$hler potential for all three chiral superfields. The
$N=1$ supergravity has the   K$\ddot{a}$hler potential and superpotential
given by 
\beqn   
K&=&\frac{\vert S\vert^2+\vert\Phi_+\vert^2+\vert\Phi_-\vert^2}{M_p^2}
\quad , \nonumber\\
W &=& \sqrt{2}gS(\Phi_+\Phi_- -\xi_+/2)\quad .\label{eq-5}
\eeqn
Using the superpotential
\beq \label{eq-6} 
W=S\frac{\partial W}{\partial S}\quad , \quad KW=\frac{S\bar{S}}
{M_p^2} \frac {\partial W}{ \partial S}
\eeq
the effective potential becomes
\beqn \label{eq-7} 
V &  &= 2g^2e^{\vert S\vert^2/M_p^2}\{\vert\frac{\partial W}{\partial
S}\vert^2( (1+\frac{S\bar{S}}{M_p^2})^2-3\frac{S\bar{S}}{M_p^2}) \no \\ 
 + && \vert S\Phi_+\vert^2+\vert S\Phi_-\vert^2\}+ \frac{g^2}{2}(\vert
\Phi_+\vert^2-\vert\Phi_-\vert^2-\xi_3)^2. 
\eeqn
After the tree level supergravity corrections, the complete potential reads
\beqn  
V &  &= 2g^2e^{\vert S\vert^2/M_p^2}\{\vert\Phi_+\Phi_-
-\xi_+/2\vert^2(1-\frac{S\bar{S}}{M_p^2}+(\frac{S\bar{S}}{M_p^2})^2)\no \\
+&  & \vert S\Phi_+\vert^2+\vert S\Phi_-\vert^2\}+
\frac{g^2}{2}(\vert\Phi_+\vert^2-\vert\Phi_-\vert^2-\xi_3)^2\; .\label{eq-8} 
\eeqn
Adding the one-loop gauge theory corrections and considering that the
inflation takes place at $\Phi_+=\Phi_-=0$, we find that the effective
potential is given by the expression
\beq  \label{eq-9} 
V=\frac{g^2\xi^2}{2}[1+\frac{g^2}{8\pi}\ln\frac{\vert S^2\vert}{\vert
S_e^2\vert}+f\frac{\vert S^4\vert}{2M_p^2}+...] \quad .
\eeq
where $f=\frac{\xi_1^2+\xi_2^2}{\xi^2}$ and $0<f<1$.

The point $\Phi_+=\Phi_-=0$ should be stable during the slow roll period in
order that the argument here presented be correct. That this is true
follows from the fact that the second derivative of (\ref{eq-8}) with
respect to  $\Phi_+$ and  $\Phi_-$ is positive as long as $2{\rm e}
^{S^2/M_p^2} S^2 > \xi_3$, which is necessary but can be achieved, as we
also see {\it a posteriori}.

Switching to the canonically quantized fields
$\phi_{\pm}=\sqrt{2}\Phi_{\pm}$ and $s=\sqrt{2}S$, the effective potential
in units $M_p=1$ is 
\beq   \label{eq-10}  
V=\frac{g^2\xi^2}{2}[1+\frac{g^2}{8\pi}\ln\frac{\vert s^2\vert}{\vert
s_e^2\vert}+f\frac{\vert s\vert^4}{8}+...]\quad .
\eeq
A general P-term inflation model has $0<f<1$ with the special case $f=0$
corresponding to the D-term inflation, while $f=1$ corresponds to the
F-term inflation. Above, $s_e$ is the bifurcation point indicating the end
of inflation. The second term in the potential is due to the one-loop
correction and the third term to the supergravity correction. 

\begin{figure}[t]
\includegraphics[width=6cm,height=5cm,angle=0]{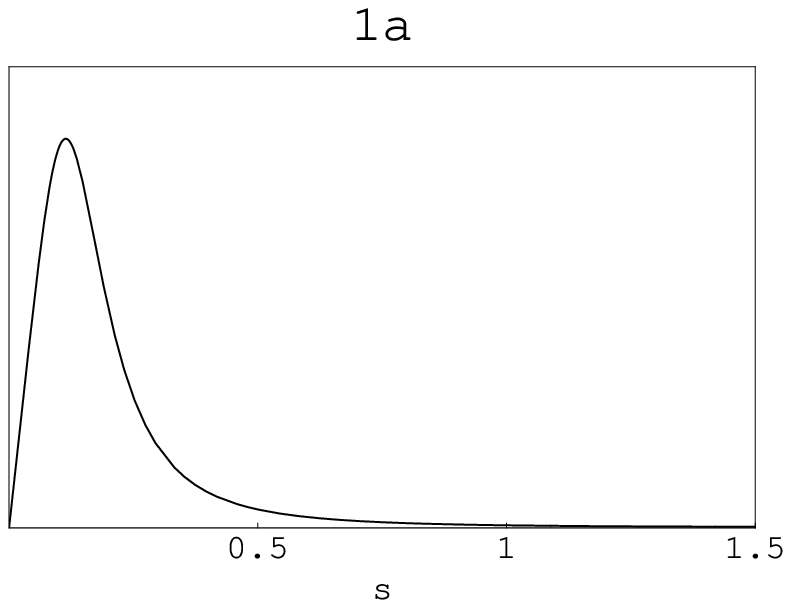}
\includegraphics[width=6cm,height=5cm,angle=0]{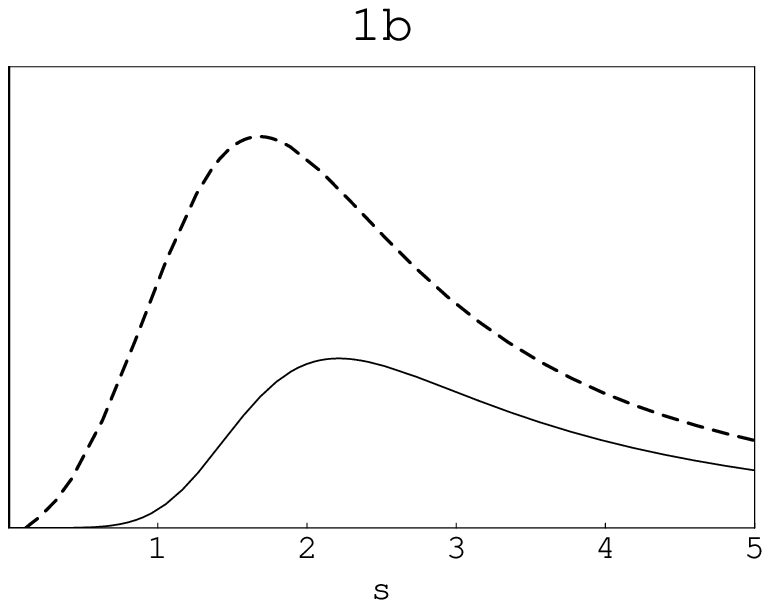}\\
\vspace{0.7cm}
\includegraphics[width=6cm,height=5cm,angle=0]{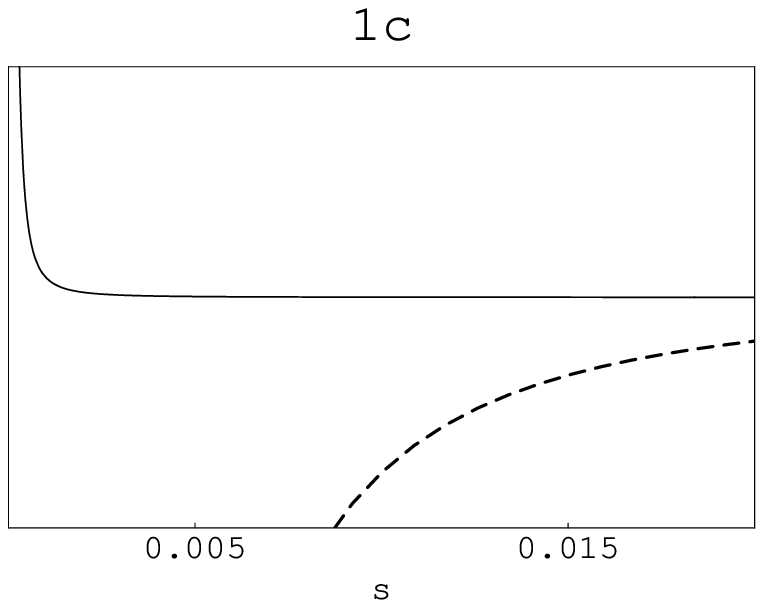}
\caption{Fig. 1a indicates the behavior of $V/V'$. Figs. 1b and 1c show
  the behavior of slow-roll parameters for large and small scalar fields,
  respectively. Dashed line indicates the $\eta$ and the solid line is
  $\epsilon$.} \label{f1} 
\end{figure}
\begin{figure}[t]
\includegraphics[width=6.7cm,height=5.58cm,angle=0]{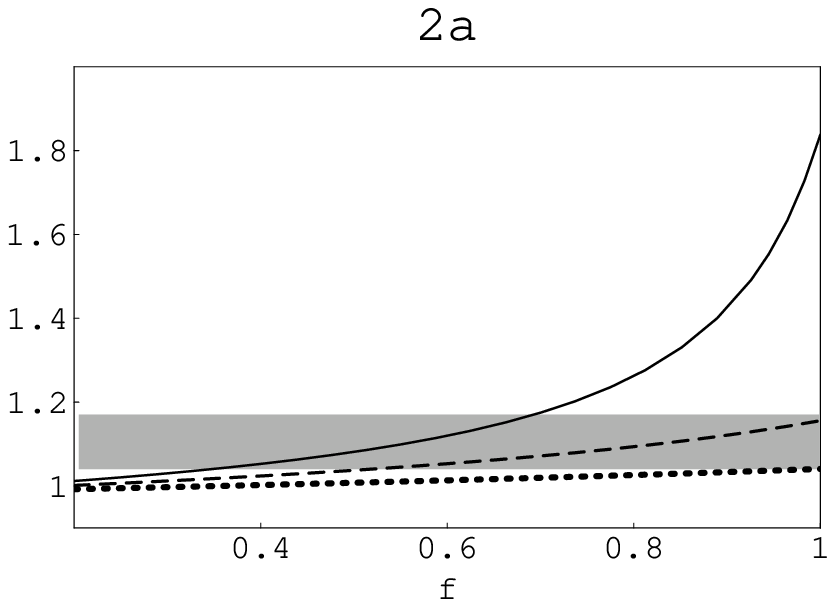}
\includegraphics[width=6.7cm,height=5.58cm,angle=0]{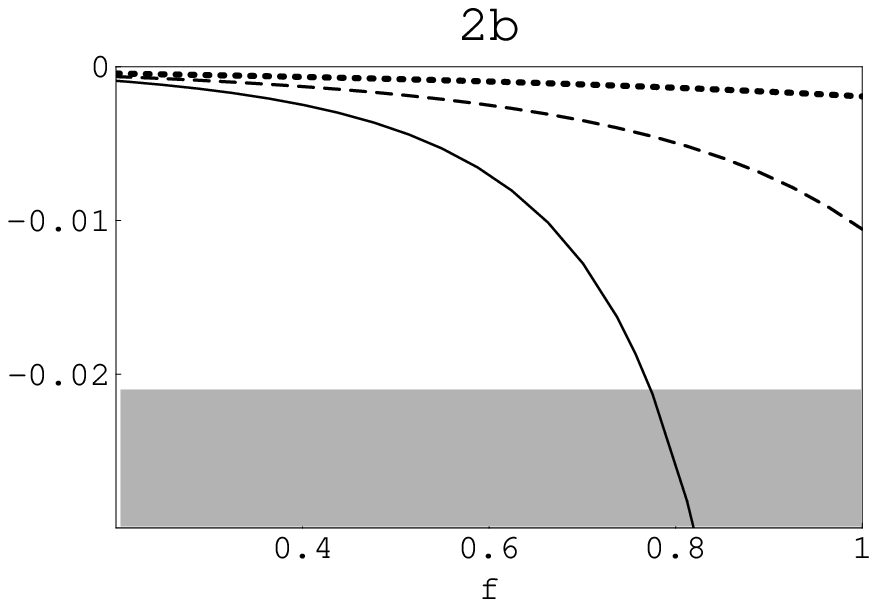} 
\caption{Dependence of the spectral index and its logarithmic derivative
  on $f$ for different fixed values of $g$ when $N_k=60$. Fig. 2a depicts
  $n$ and Fig.2b $dn/d\ln k$. A solid line corresponds to $g=0.11$, a
  dashed line to $g=0.09$, and a dotted line to $g=0.065$.} \label{f2}
\end{figure}
\begin{figure}[t]
\includegraphics[width=6cm,height=5cm,angle=0]{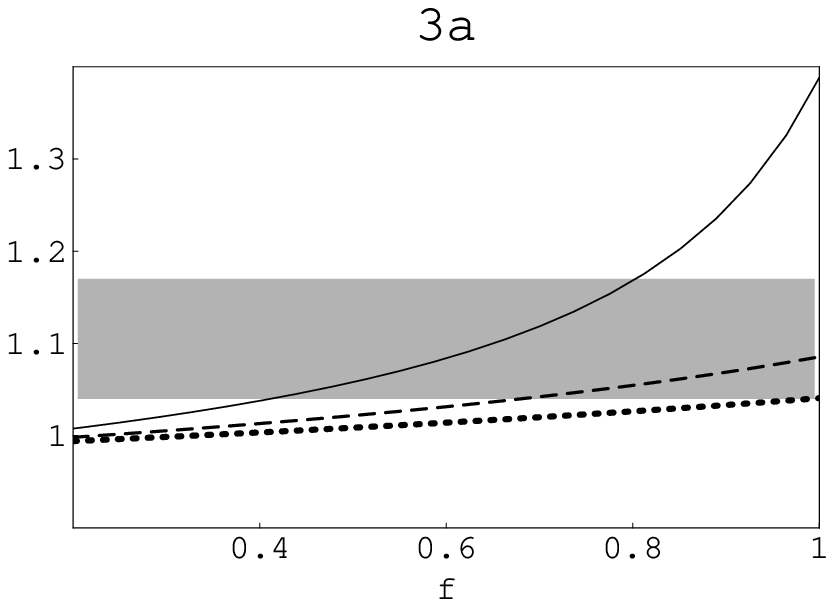}
\includegraphics[width=6cm,height=5cm,angle=0]{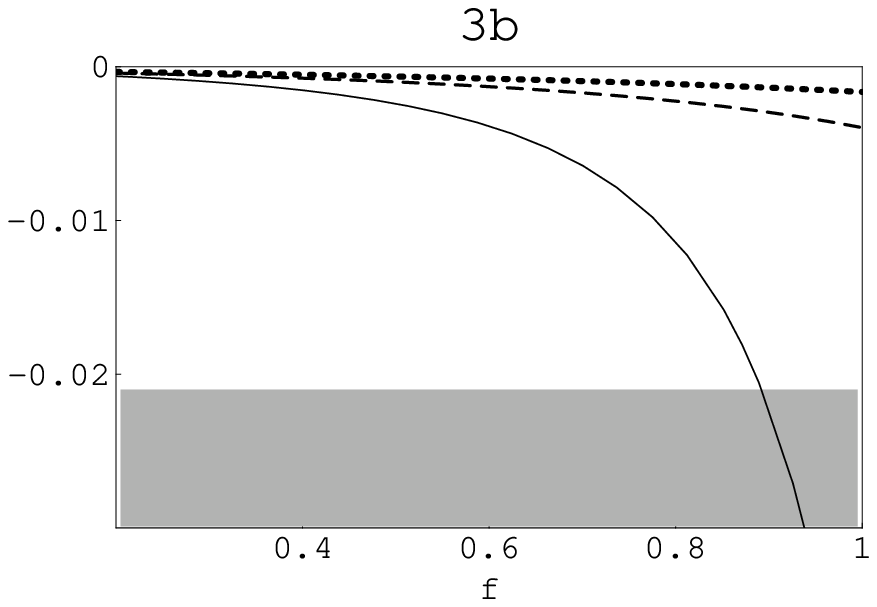}\\
\caption{Dependence of the spectral index and its logarithmic derivative
  on $f$ for different fixed $g$ when $N_k=70$.  Fig. 3a shows $n$ while
  Fig. 3b shows $dn/d\ln k$. A solid line corresponds to $g=0.09$, a
  dashed line to $g=0.07$ and a dotted line to $g=0.058$. } \label{f1-2}
\end{figure}
\begin{figure}[t]
\includegraphics[width=6.7cm,height=5.58cm,angle=0]{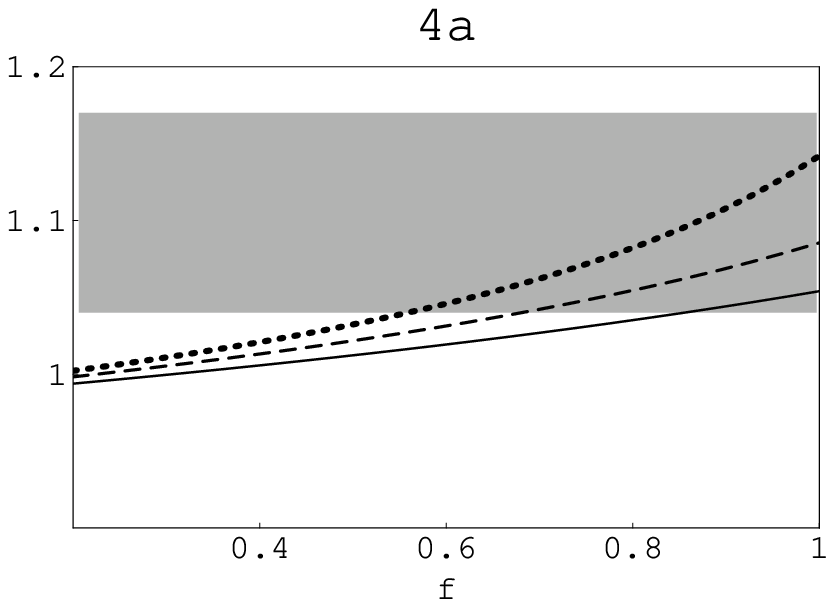}
\includegraphics[width=6.7cm,height=5.58cm,angle=0]{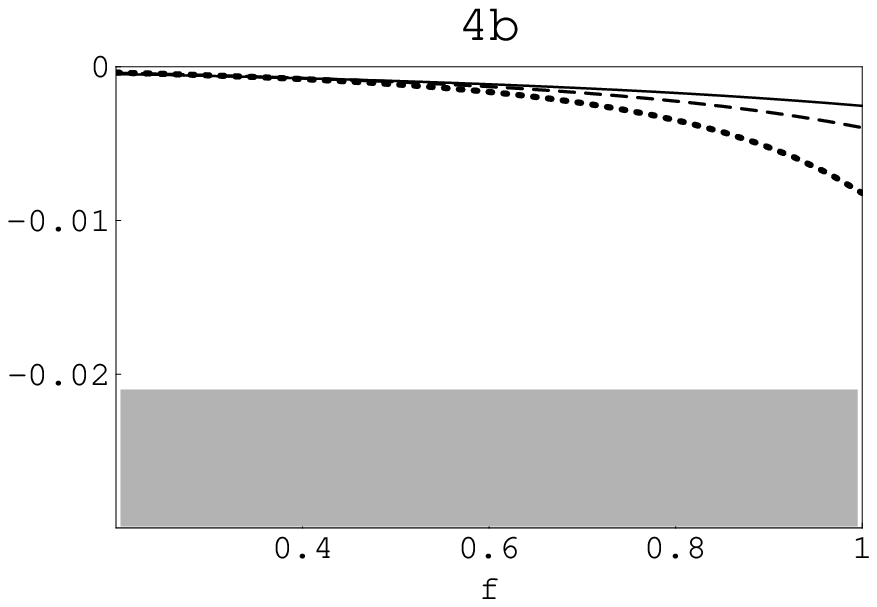} 
\caption{Dependence of the spectral index and its logarithmic derivative
  on $f$ for different numbers of e-foldings when $g$ is fixed.  Fig. 4a
  corresponds to $n$ and Fig. 4b to $dn/d\ln k$. A solid line corresponds
  to $N_k=60$, a dashed line to $N_k=70$ and a dotted line to $N_k=80$.}
  \label{f2-2} 
\end{figure}
\begin{figure}[t]
\includegraphics[width=6cm,height=5cm,angle=0]{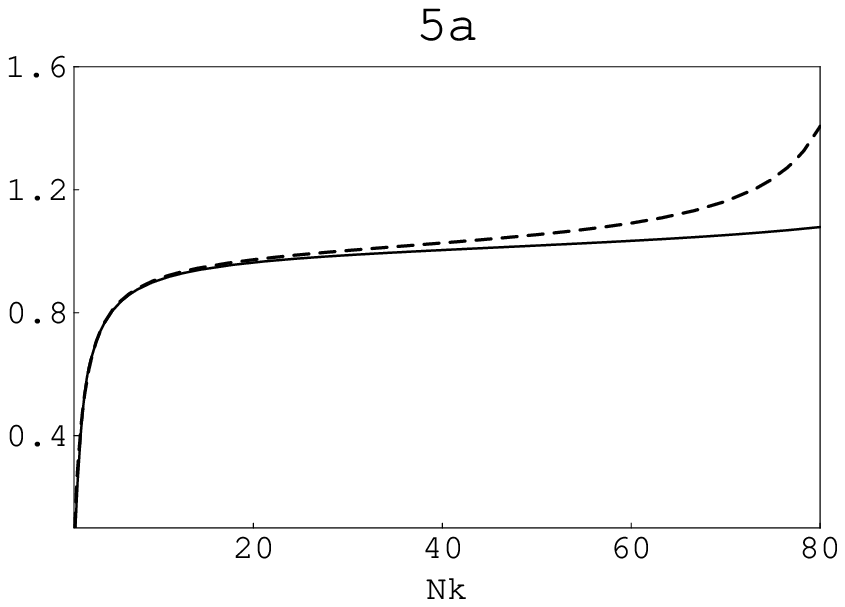}
\includegraphics[width=6cm,height=5cm,angle=0]{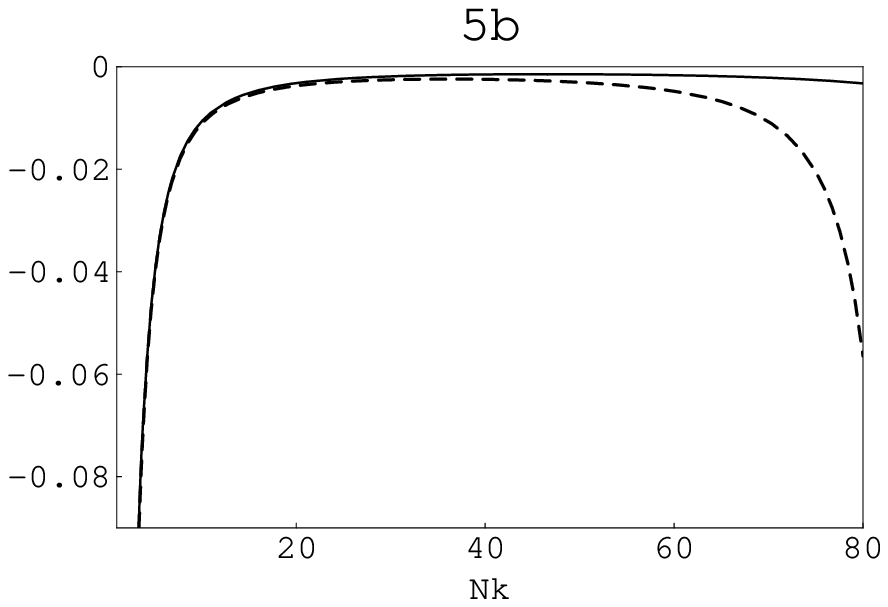}\\
\caption{Dependence of the spectral index and its logarithmic derivative
  on the number of e-foldings for fixed $f$ and $g$.  Fig. 5a describes
  $n$ and Fig. 5b  $dn/d\ln k$. A solid line corresponds $g=0.08, f=0.6$,
  a dashed line to $g=0.08, f=1$.} \label{f1-3} 
\end{figure}
\begin{figure}[t]
\includegraphics[width=6.7cm,height=5.58cm,angle=0]{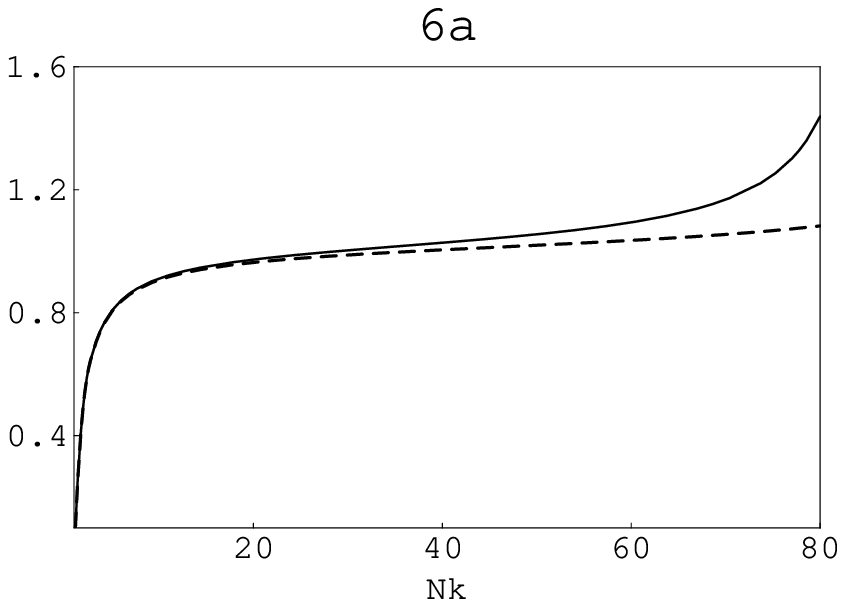}
\includegraphics[width=6.7cm,height=5.58cm,angle=0]{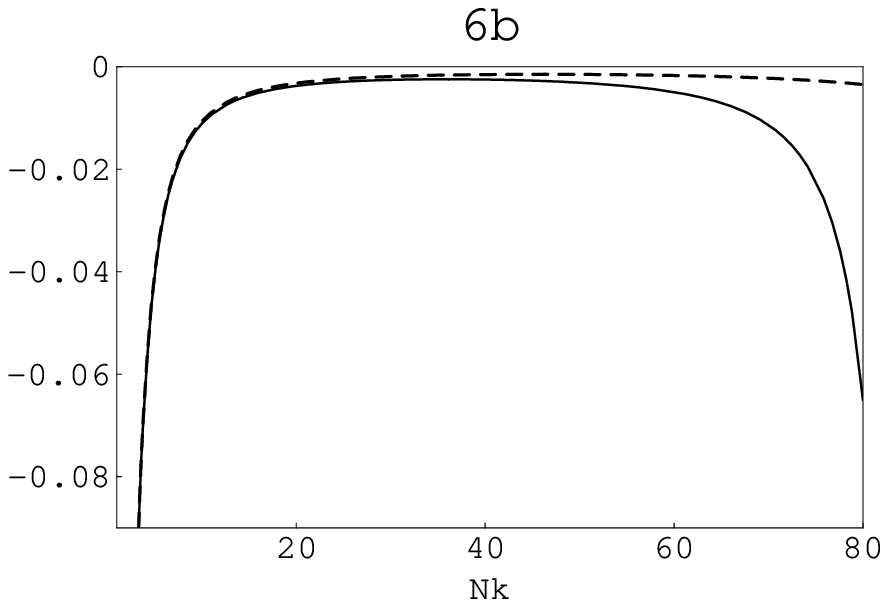} 
\caption{Dependence of the spectral index and its logarithmic derivative
  on the number of e-foldings for fixed $f$ and $g$. Fig. 6a describes $n$
  and Fig. 6b $dn/d\ln k$. A solid line corresponds to $f=0.8, g=0.09$, a
  dashed line to $f=0.8, g=0.07$.} \label{f1-4} 
\end{figure}

\section{The inflationary space
\label{sec:3}}
In a single field slow-roll inflation model with a potential $V(s)$ the
amplitude of curvature perturbation is given by 
\beq    
P(k)=\frac{1}{2\pi}\frac{H^2(t_k)}{\vert \dot{s}(t_k)\vert},
\eeq\label{eq-11}
where
\beq   
H^2(t_k)=V[s(t_k)]/3.
\eeq\label{eq-12}
and $t_k$ is the epoch where the $k$ mode left the Hubble radius during
inflation \cite{1}. As usual, the spectral index is defined by
\beq   
n-1=\frac{d\ln\vert P(k)\vert^2}{d\ln k}=-6\epsilon+2\eta,
\eeq\label{eq-13}
where 
\beq  \label{eq-14} 
\epsilon=\frac{1}{2}[V'(s)/V(s)]^2\quad , \quad \hbox{and}\quad  
\eta=V''(s)/V(s).
\eeq
The logarithmic derivative of the spectral index is 
\beq  \label{eq-15} 
\frac{d n}{d\ln k}=16\epsilon\eta-24\epsilon^2-2\zeta,
\eeq
where
\beq  \label{eq-16} 
\zeta=\frac{V'''(s)V'(s)}{V^2(s)}.
\eeq
It has been reported that WMAP results favors purely adiabatic
fluctuations with a remarkable feature that the spectral index runs from
$n>1$ on a large scale to $n<1$ on a small scale. More specifically on the
scale $k=0.002Mpc^{-1}, n=1.10^{+0.07}_{-0.06}$ and $dn/d\ln
k=-0.042^{+0.021}_{-0.020}$ \cite{5}. It is of interest to investigate
whether the P-term inflation can accommodate these observational result. 

From (\ref{eq-12}), for the extremely flat potential, the scale factor of
the universe is
\beq  \label{eq-17}  
a(t)=a(0)exp(g\xi t/\sqrt{6}),
\eeq
and the approximate number of e-foldings can be written as $N=g\xi t/\sqrt{6}$.

From the potential form (\ref{eq-10}) we learnt that inflation consists of
two long stages, one of them is determined by the one-loop effect and the 
other is determined by the supergravity corrections. Comparing the
derivatives of the one-loop and supergravity correction terms in the
potential ((\ref{eq-10}), we learnt that if $V'_{1-loop}>V'_{sugra}$, then
$\vert s\vert<\sqrt{g}/(2\pi^2)^{1/2}$. Therefore one gets $N_1\sim 4.5/(f^
{1/2}g) $. Similarly for the $V'_{sugra}>V'_{1-loop}$, one has $N_2\sim 4.5/
(f^{1/2}g)$. The total duration of inflation can be estimated by
\beq   \label{eq-18}   
N_1+N_2\sim 9/(f^{1/2}g)>N_k,
\eeq
where $N_k$ is supposed to be a reasonable number of e-foldings. 

Thus we require $g<9/(N_k f^{1/2})$. For the F-term inflation $f=1$ and
$N_k=60$, $g<0.15$, which is exactly the argument given in \cite{7}. 

Using (\ref{eq-10}) and (\ref{eq-14},\ref{eq-16}), we have for the 
P-term inflation
\beqn   
\epsilon & = &\frac{1}{2}[\frac{g^2/(4\pi^2\vert s\vert)+f\vert
s\vert^3/2}{1+g^2/(8\pi^2)\ln(\vert s\vert^2/\vert s_e\vert^2)+f\vert
s^4\vert/8}]^2 \quad ,\no \\
\eta & = & \frac{-g^2/(4\pi^2\vert s\vert^2)+3f\vert
s\vert^2/2}{1+g^2/(8\pi^2)\ln(\vert s\vert^2/\vert s_e\vert^2)+f\vert
s\vert^4/8}\quad , \label{eq-19}\\ 
\zeta & = & \frac{[g^2/(2\pi^2\vert s\vert)+f\vert
s\vert^3][g^2/(4\pi^2\vert s\vert^3)+3f\vert
s\vert/2]}{(1+g^2/(8\pi^2)\ln(\vert s\vert^2/\vert s_e\vert^2)+f\vert
s\vert^4/8)^2}\quad . \no 
\eeqn
The number of e-foldings during inflation, is given by \cite{1}
\beqn  
&&N(s) = -\int^{s_{end}}_s V/V' ds \no \\
& = &\int^s_{s_{end}}\frac{1+g^2/(8\pi^2)\ln(\vert s\vert^2/\vert
s_e\vert^2)+f\vert s\vert^4/8}{g^2/(4\pi^2\vert s\vert)+f\vert
s\vert^3/2}, \label{eq-20}
\eeqn
where $V/V'$ has the behavior shown in Fig.1a with a maximum value when
$s=s_0$.  

The value $s_{end}$ indicates the end of inflation, which can be obtained from the condition
\beq  \label{eq-21} 
max\{\epsilon(s_{end}), \eta(s_{end})\}=1\quad ,
\eeq
when $\eta$ tends to unit first, its $s_{end}$ is always much larger than
$s_0$ as shown in Fig.1b, where the dashed line indicates the behavior of
$\eta$ for large $s$. The integral (\ref{eq-20}) calculated using the
value of $s_{end}$ determined by $\eta=1$ fails to give a reasonably large
values for the number of e-foldings to solve the horizon and the entropy
problems, as required by inflation. Thus, such an $s_{end}$ is not the
real end point of inflation. For small values of $s$, $\eta$ becomes
negative (see dashed line in Fig.1c). The behavior of $\epsilon$ is shown
in solid lines in Fig.1b and Fig.1c for large $s$ and small $s$
respectively. For large $s$, as the case of $\eta$, the value of $s_{end}$
gotten from $\epsilon=1$ is not useful. Enough number of e-foldings can
only be obtained from the integral (\ref{eq-20}) by using the value of
$s_{end}$ obtained from $\epsilon=1$ for small $s$ ($s_{end}<s_0$). This
value of $s_{end}$ is the real end point of inflation. 

Now we adopt the following strategy applied in \cite{12}. From
(\ref{eq-20}) we can express $s (s=s_k)$ as a function of $s_{end}$ and
$N_k$ for different values of $f$ and $g$. $N_k$ is the number of
e-foldings between the time the scales of interest leave the horizon and
the end of inflation. Inserting such an $s$ into (\ref{eq-19}) and using
(\ref{eq-13}) and (\ref{eq-15}) we can obtain the spectral index and its
logarithmic derivative. The numerical results and the comparison with the
WMAP observation are below. 

For $f=0$, which corresponds to the D-term inflation, our numerical
calculation gives $n=1$ and $dn/d\ln k=0$, which is in exact in agreement
with the argument in \cite{9}. 

For a general P-term inflation model with arbitrary $0<f<1$, we found a
richer physics. Figs. 2 and 3 show the dependence of the spectral index
and its logarithmic derivative on $f$ for different values of $g$ when the
number of e-foldings are 60 and 70, respectively. The shadows indicate the
WMAP observational range. We learnt that there is a threshold value of
$g_{min}$ to force the spectral index $n$ to meet the minimum
observational result 1.04, $g_{min}=0.065$ for $N=60$ and $g_{min}=0.058$
for $N=70$. Smaller values of $g_{min}$ are ruled out by observation. With
the increase of $N$, the threshold value $g_{min}$ can be smaller. However
due to the existence of the upper bound of the number of e-foldings
\cite{13}, this threshold value of $g$ cannot be reduced arbitrarily. 
Consequently the attempt to suppress the cosmic string contribution to 
perturbations of metric by choosing a small enough gauge coupling $g\ll
1.3\times 10^{-5}$ is hampered. Indeed, cosmic strings have energy density
per unit length equal to $\mu=2\pi \xi$. In the D-term inflation, the 
inflationary perturbations on the horizon scale is
$ \delta_H \sim V^{3/2}/V' = (2\sqrt{2} \pi^2 \xi s_N) /g \sim 10^{-4}$,
where $s_N$ is the initial value of the scalar field given by the
expression
\beq
s_N^2= s_c^2 + (g^2 N)/(2\pi^2)=\xi + (g^2 N)/(2\pi^2)
\eeq
and $s_c=\sqrt{\xi}$ is the bifurcation point, while $N$ is the 
number of e-foldings when the field rolls from $s_N$ until the
bifurcation point $s_c$.

If $g$ is small, $g^2 N/(2\pi^2) << \xi$,
and $V^{3/2}/V' \approx (2\sqrt{2} \pi^2 \xi^{3/2})/g \sim 10^{-4}$
Therefore $g \sim 2\sqrt{2} \pi^2 \xi^{3/2} \times 10^{4} = \pi^{1/2}
\mu^{3/2}\times 10^{4}$, which shows that for relatively small gauge
coupling, the cosmic string energy density is related to the
gauge coupling constant. The smaller the gauge coupling constant is, the
smaller the energy density of the cosmic string will be, which leads to the
small constribution of the cosmic string to the density fluctuations. It
is naively expected that for the P-term inflation this relation is
inherited, since the only difference from the D-term inflation is the
supergravity correction. Since we have a lower limit for the gauge
coupling constant the conclusion about the importance of cosmic strings is
inevitable.

Some alternative mechanism to suppress the
contribution of strings is required \cite{9}. We see from Figs. 2 and 3,
for fixed $f$, both the values of the spectral index and its logarithmic
derivative increase with the increase of $g$. For fixed $g$, they increase
with $f$ as well. However we cannot enforce both $n$ and $dn/d\ln k$ to
meet the WMAP observational result at the same time for the common range
of $f$ and $g$. When $f=1$, the model goes to the F-term inflation. With
larger values of $g$, we can find larger values of $dn/d\ln k$ as shown in
those figures. It is not as mild as $10^{-3}$, as claimed in \cite{11}.

For fixed $g$, the dependence of the spectral index and its logarithmic
derivative on $f$ for different values of the number of e-foldings is
shown in Fig. 4. We find that with the increase of the number of
e-foldings both $n$ and $\vert dn/d\ln k\vert$ increase. 

Figs. 5 and 6 show that the spectral index and its logarithmic derivative
depend on the number of e-foldings for different fixed values of $g$ and
$f$. It is clear that with the decrease of the number of e-foldings, it is
possible to have the spectral index logarithmic derivative from $n>1$ to
$n<1$. Considering $exp(N)=k_e/k$, where $k_e$ indicates the scale at the
end of inflation, this result is consistent with the WMAP behavior, {\it
i. e.} $n$ decreases from $n>1$ to $n<1$ as $k$ increases. Furthermore,
Figs. 5 and 6 tell us that at the beginning of the inflation, when $N(k)$
is small, the spectral index changes quickly, and its logarithmic
derivative is larger. When the number of e-foldings is large enough, as
{\it e. g.}  over 75 for $f=0.7, g=0.09$, the logarithmic derivative of
the spectral index can meet the observational requirement ($-0.021\sim
-0.062$), however the spectral index will be bigger than the observational
range. Again, $n$ and $\vert dn/d\ln k\vert$ cannot comply with 
observation at the same time.

\section{Conclusion
\label{sec:4}}
We have analysed the implications of WMAP results, in particular the
bounds on the inflation observables, for the P-term inflationary model. We
found that compared to the D-term or the F-term inflation alone, the
P-term inflation model with a running parameter $0<f<1$ displays a richer
physics. In addition to the upper bound on $g$ determined by a reasonable
number of e-foldings to solve the horizon problem as required by the
inflation, the observational data of the spectral index together with the
upper limit of the number of e-foldings puts the lower bound on the choice
of $g$. This lower bound on $g$ makes it not possible to suppress the
cosmic string contribution to the inflationary perturbations by counting
on simply choosing too small values of $g$. It gives more motivation to
call for other alternative mechanism to suppress the string contribution
\cite{9}. With the changable parameter $f (0<f<1)$, we have observed the
possiblity of obtaining a logarithmic derivative spectral index such that
$n>1$ on large scales while $n<1$ on small scale for P-term inflation,
which is consistent with WMAP observation. The dependence on the spectral
index and its logarithmic derivative on the model parameters $g$ and $f$
has been investiged. We have also studied whether it is possible to obtain
a spectral index and its logarithmic derivative within the observational
range for certain $f$ and $g$ and concluded that it is not possible to
accommodate both observational ranges of $n$ and $dn/d\ln k$ at the same
time. When the spectral index is within the observational range
$(1.04-1.17)$, its logarithmic derivative is always smaller than the WMAP
data $(-0.021\sim -0.062)$, however it is not as mild as claimed in
\cite{11}, when $f=1$. The larger values of the logarithmic derivative of
the spectral index can be around $-0.011$ for values of $f$ and $g$
keeping the spectral index $n$ within the WMAP range. This behavior is
qualitatively consistent with most existing inflationary models satisfying
slow roll condition. With the tunable parameters $f$ and $g$ in the P-term
inflation, the logarithmic derivative of the spectral index can be
increased, which shows the benefit of this model. 

ACKNOWLEDGEMENT: This work was partially supported by FAPESP and CNPQ,
Brazil. B. Wang would like to acknowledge the support by NNSF, China,
Ministry of Education of China and  Ministry of Science and Technology of
China under Grant NKBRSFG19990754. The work of Chi-Yong Lin was supported
in part by the National Science Council under the Grant
NSC-92-2112-M-259-009. 



\begin{thebibliography}{nn}
\bibitem{1} D. H. Lyth and A. Riotto, Phys. Rep. {\bf 314}, 1 (1999).
\bibitem{2} A. D. Linde, Particle Physics and Inflation Cosmology, Harwood
Academic, Swizerland (1990). E. W. Kolb and M. S. Turner, The Early
Universe, Addison Wesley (1990).  
\bibitem{3} E. L. Wright {\it et al}, Astrophys. J. {\bf 420} 1 (1994).
\bibitem{4} E. Torbet {\it et al}, Astrophys. J. {\bf 521} L79 (1999);
P. D. Manskopf {\it et al}, Astrophys. J. {\bf 536} L59 (2000); A. Beneit,
astro-ph/0210305. 
\bibitem{5} C. L. Bennett {\it et al}, astro-ph/0302207; D. N. Spergel
{\it et al.}, astro-ph/0302209; H. V. Peiris {\it et al.},
astro-ph/0302225. 
\bibitem{6} P. Binetruy and G. Dvali, Phys. Lett. B {\bf 388}, 241 (1996);
E. Helyo, Phys. Lett. B {\bf 387}, 43 (1996). 
\bibitem{7} E. J. Copeland, A. R. Liddle, D. H. Lyth, E. D. Stewart and
D. Wands, Phys. Rev. D {\bf 49}, 6410 (1994); G.  Dvali, Q. Shafi and
R. Schaefer, Phys. Rev. Lett. {\bf 73}, 1886 (1994). 
\bibitem{8} A.  Linde and A. Riotto, Phys. Rev. D {\bf 56}, 1841 (1997).  
\bibitem{9} R. Kallosh and A. Linde, hep-th/0306058; R. Kallosh,
hep-th/0109168.  
\bibitem{kinney} W. H. Kinney, E. W. Kolb, A. Melchiorri and A. Riotto,
hep-th/0305130; B. Feng, M. Li, R. J. Zhang and X. Zhang,
astro-ph/0302479; M. Bastero-Gil, K. Freese and L. M. Houghton,
hep-ph/0306289; B. Feng and X. Zhang, astro-ph/0305020; 
Arman Shafieloo, Tarun Souradeep astro-ph/0312174; Ryo Nagata, Takeshi
Chiba, Naoshi Sugiyama astro-ph/0311274; Bo Feng, Mingzhe Li, Ren-Jie
Zhang, Xinmin Zhang {\it Phys. Rev.} {\bf  D68} (2003) 103511;
Gia Dvali, Shamit Kachru hep-ph/0310244; Utpal Chattopadhyay , Achille
Corsetti, Pran Nath {\it Phys.Rev.} {\bf D68} (2003) 035005; John
R. Ellis, Keith A. Olive, Yudi Santoso, Vassilis C. Spanos,  
{\it Phys. Lett.} {\it B565} (2003) 176; Jean-Philippe Uzan, Ulrich
Kirchner, George F.R. Ellis  {\it Mon. Not. Roy. Astron. Soc. } {\bf L65} 
344.
\bibitem{10} B. Kyae and Q. Shafi, astro-ph/0302504.
\bibitem{11} M. Kawasaki, M. Yamaguchi and J. Yokoyama, Phys. Rev. D {\bf
68}, 023508 (2003). 
\bibitem{12} M. Bento, N. Santos and A. Sen, astro-ph/0307093;
astro-ph/0307292. 
\bibitem{13} S. Dodelson and L. Hui, {\it Phys. Rev. Lett.} {\bf 91}
(2003) 131301 astro-ph/0305113; A. Liddle and
S. Leach, astro-ph/0305263; T. Banks and W. Fischler, astro-ph/0307459;
B. Wang and E. Abdalla, hep-th/0308145. 
\end{thebibliography}
\end{document}